\documentstyle[subfigure,psfig,floats,aps,aipbook]{revtex}

\psfigurepath{/home/bulls/tomek/orbits/paper}

\def\kms{\hbox{km~s$^{-1}$}}
\def\mathnew{\mathsurround=0pt}
\def\simov#1#2{\lower .5pt\vbox{\baselineskip0pt \lineskip-.5pt
        \ialign{$\mathnew#1\hfil##\hfil$\crcr#2\crcr\sim\crcr}}}
\def\simgreat{\mathrel{\mathpalette\simov >}}
\def\simless{\mathrel{\mathpalette\simov <}}

\begin{document}
\title{Gamma-Ray Bursts from High Velocity Neutron Stars}
\author{Tomasz Bulik and Donald Q. Lamb}
\address{Department of Astronomy and Astrophysics\\
University of Chicago\\
5640 South Ellis Avenue\\
Chicago, IL 60637}
\maketitle
\begin{abstract}
We investigate the viability of the Galactic halo model
of $\gamma$-ray bursts by calculating the spatial distribution
of neutron stars born with high velocities in the Galactic
disk, and comparing the resulting brightness and angular distribution
with the BATSE data.
We find that the Galactic halo model can reproduce the BATSE
peak flux and angular distribution data for neutron star kick
velocities $\simgreat 800$~\kms, source turn-on ages $\simgreat 10$~Myrs,
and sampling depths $100{\rm~kpc}\simless d_{max} \simless
400$~kpc.
\end{abstract}
%


\section*{Introduction}

Gamma-ray bursts (GRBs) continue to confound astrophysicists nearly a
quarter century after their discovery \cite{KSO:739}.  Before the
launch of CGRO, most scientists thought that GRBs came from magnetic
neutron stars residing in a thick disk (having a scale height of up to
$\sim$ 2 kpc) in the Milky Way \cite{Hig:Lin:90,Hard:91}.  The data
gathered by BATSE showed the existence of a rollover in the cumulative
brightness distribution of GRBs and that the sky distribution of even
faint GRBs is consistent with isotropy \cite{Meegan:92,Briggs:95}.
This rules out a thick Galactic disk source population.

Consequently, the primary impact of the BATSE results has been to
intensify debate about whether the bursts are Galactic or cosmological
in origin.  Galactic models attribute the bursts primarily to
high-velocity neutron stars in an extended Galactic halo, which must
reach one fourth or more of the distance to M31 ($d_{\rm M31} \sim 690$
kpc) in order to avoid any discernible anisotropy
\cite{Hak:94,Hartmann:94}.  Cosmological models place the GRB sources
at distances $d \sim 1 - 3$ Gpc, corresponding to redshifts $z\sim 0.3
- 1$.  A source population at such large distances naturally produces
an isotropic distribution of bursts on the sky, and the expansion of
the universe or source evolution can reproduce the observed rollover in
the cumulative brightness distribution \cite{Fenimore:93}.

Recent studies \cite{LyneLori:94,Frail:94} have revolutionized our
understanding of the birth velocities of radio pulsars.  They show that
a substantial fraction of neutron stars have velocities that are high
enough to produce an extended halo around the Milky Way like that
required by Galactic halo models of GRBs \cite{LiDer:92}.

Podsiadlowski, Rees, and Ruderman \cite{Pods:94} have carried out
pioneering calculations of the spatial distribution expected for
high-velocity neutron stars born in the Galactic disk.  They consider
the effects of a non-spherical halo potential and of M31, but neglect
the effects of the Galactic disk, which we find is also important.

\section*{Models}
We have calculated detailed models of the spatial distribution expected
for a population of high-velocity neutron stars born in the Galactic
disk and moving in a Galactic potential that includes the bulge, disk,
and a dark matter halo.

We use the mass distribution and potential given by Kuijken and Gilmore
\cite{KG:89} which includes a disk, a bulge, and a dark matter halo.
The densities of the disk and of the halo are
\begin{eqnarray}
\rho_D= \rho_D^0 \exp\left({-r\over r_d}\right)\exp\left({-z\over z_d}\right),
&
{}~~~~~~ &
\rho_H=\rho_H^0 \left[ 1 + \left({r\over r_c}\right)^2 \right]^{-1}.
\end{eqnarray}
The circular velocity $v_c$ and the Galactic disk lead to
characteristic angular anisotropies as a function of burst brightness
which provide a signature, and therefore a test, of high-velocity
neutron star models.  Prolate or oblate dark matter halos also produce
other angular anisotropies as a function of burst brightness which may
provide a signature of such models \cite{Pods:94}.

We assume that neutron stars are born with the circular velocity $v_c
\approx 220~\kms$ of the Galactic disk.  Given that current knowledge
of the distribution of initial kick velocities is uncertain, we adopt a
Green-function approach: we calculate the spatial distribution of
neutron stars for a set of kick velocities (e.g., $v_{\rm kick} = 200,
400,..., 1400$~\kms).  We follow the resulting orbits for up to $3
\times 10^9$ years.

In our initial calculations, we assume that the bursts are standard
candles, i.e. $L = \delta(L-L_0)$.  We parameterize the burst-active
phase by a turn-on age $\delta t$ and a duration $\Delta t$, and assume
that the rate of bursting is constant throughout the burst-active
phase.  The high-velocity neutron star model then has four parameters:
$v_{\rm kick}$, $\delta t$, $\Delta t$, and the BATSE sampling depth
$d_{\rm max}$.

\begin{figure}[th]
\begin{center}
\begin{tabular}{lr}
{\psfig{file=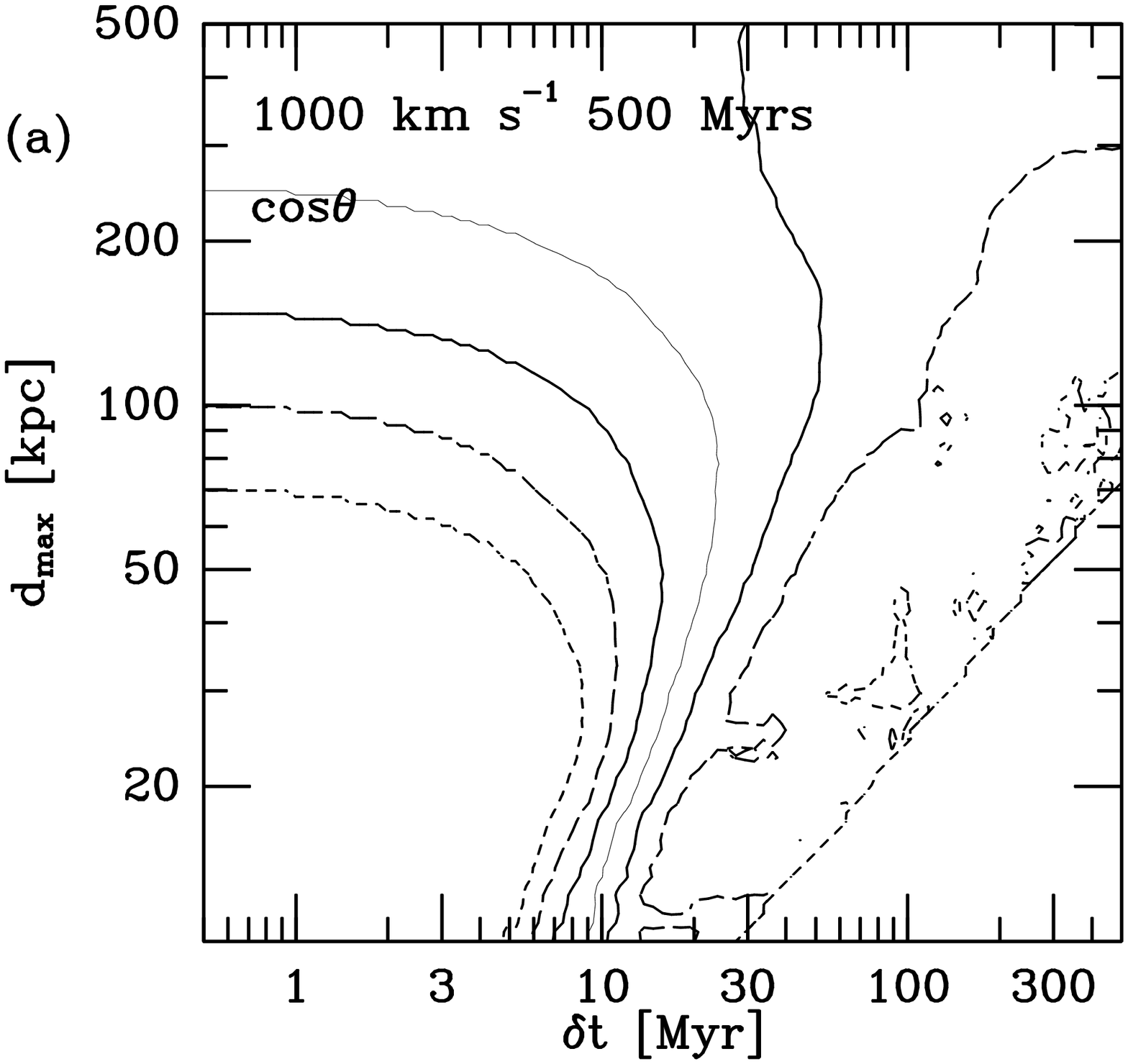,width=5.cm,angle=-90}} &
{\psfig{file=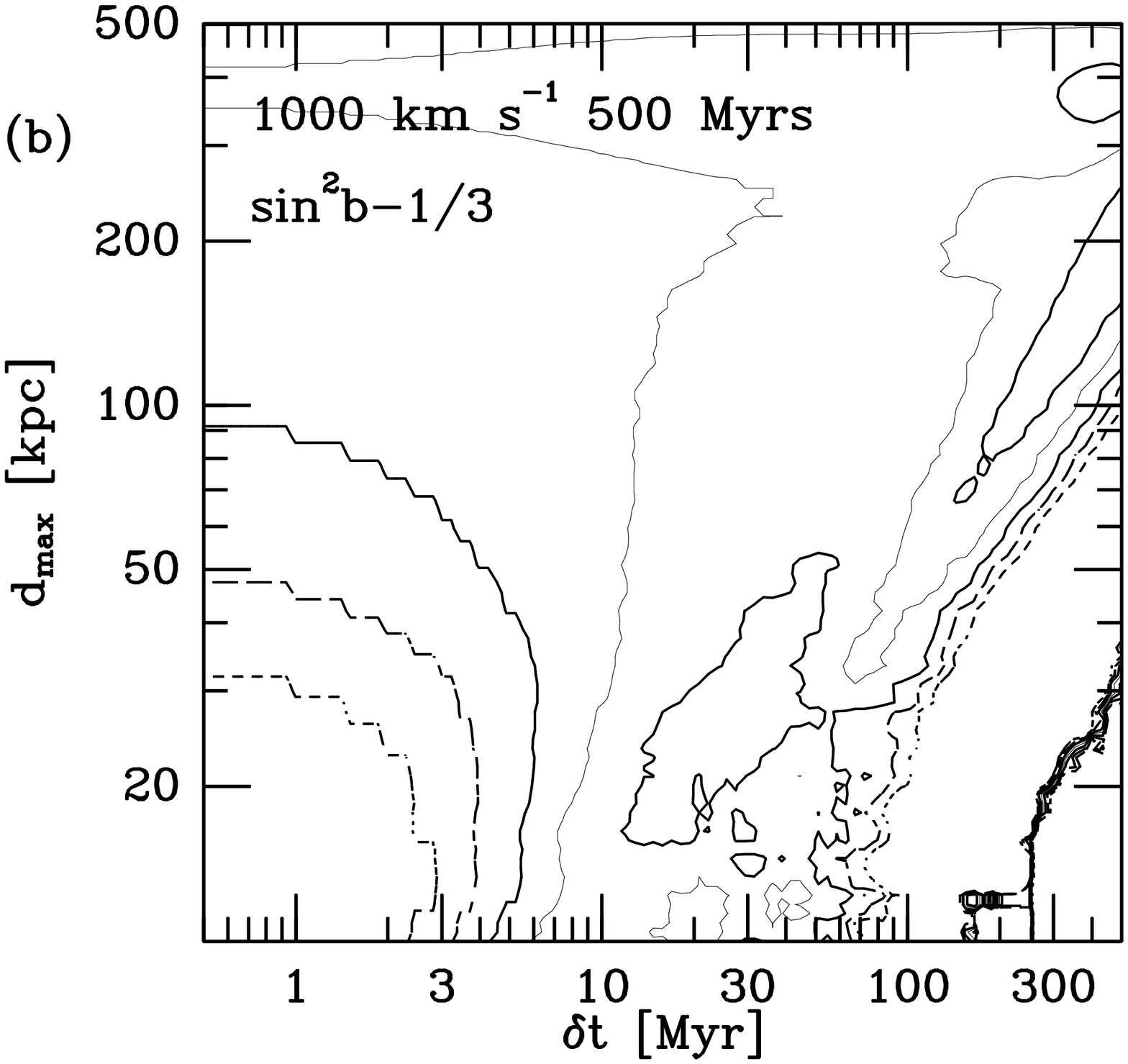,width=5.cm,angle=-90}} \\
{\psfig{file=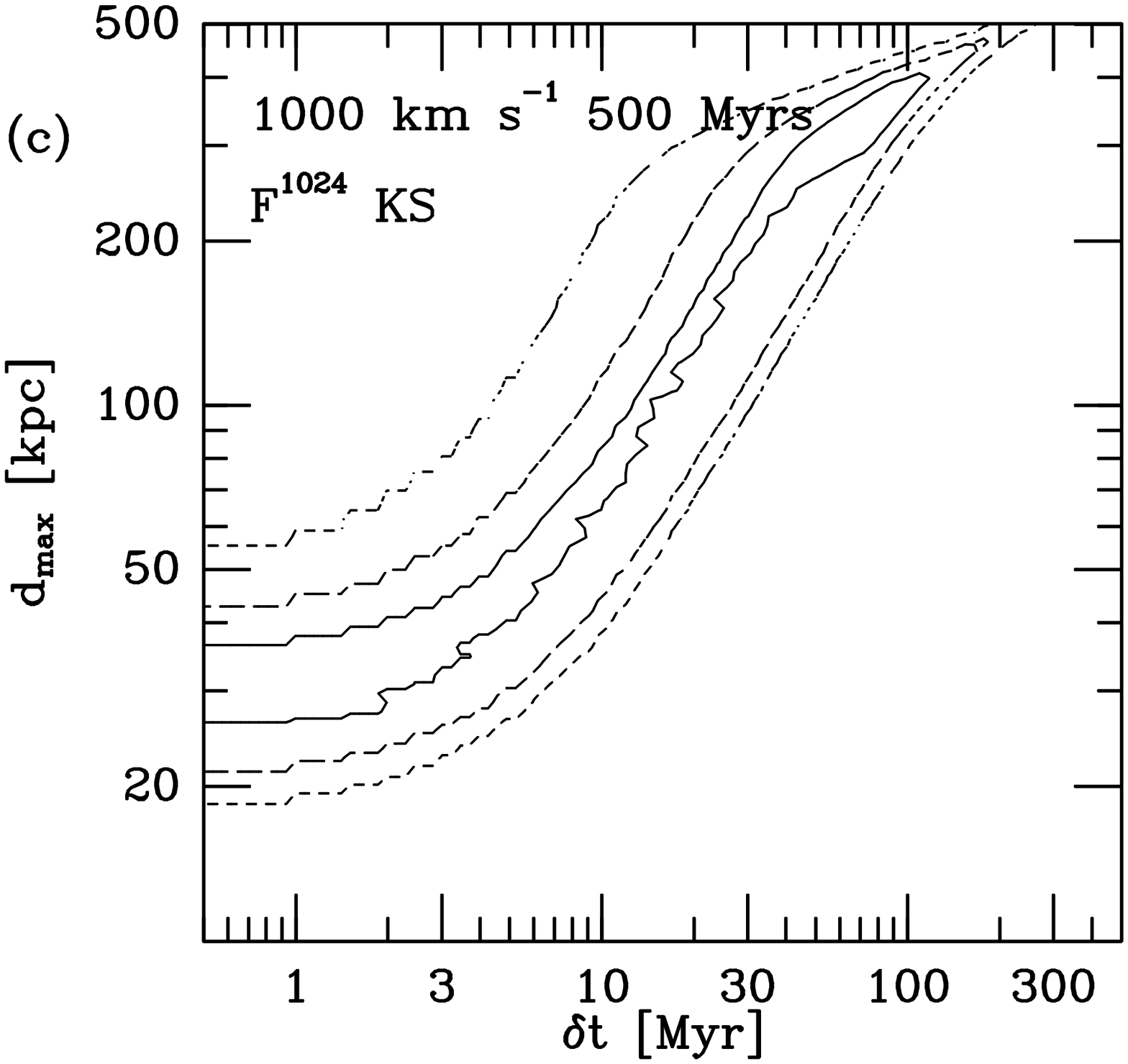,width=5.cm,angle=-90}}  &
{\psfig{file=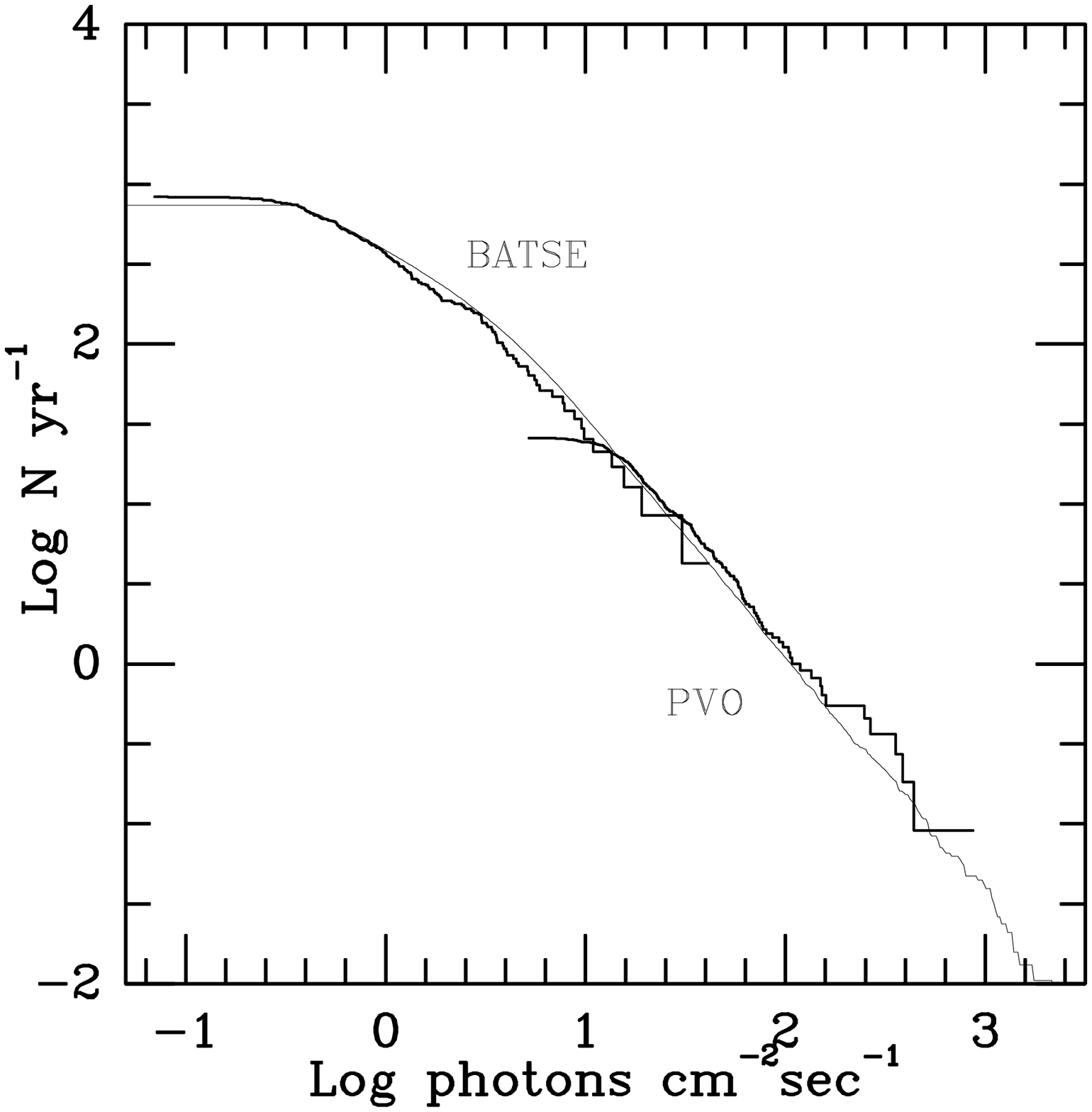,width=5.cm,angle=-90}} \\
\end{tabular}
\end{center}
\caption{Comparison of a Galactic halo model in which neutron stars
are born with a kick velocity of
$1000$~\kms\ and have a burst-active phase lasting
$\Delta t = 500$ million years with a carefully-selected sample of 285
bursts from the BATSE 2B catalogue.  Panels (a) and (b) show the
contours in the ($\delta t$, $d_{\rm max}$)-plane along which the
Galactic dipole and quadrupole moments of the model differ from those
of the data by $\pm$ 1$\sigma$ (solid lines), $\pm$ 2$\sigma$ (dashed
line), and $\pm$ 3$\sigma$ (short-dashed line) where $\sigma$ is the
model variance; the thin line in panel (a) shows the contour where the
dipole moment for the model equals that for the data.  Panel (c) shows
the contours in the ($\delta t$, $d_{\rm max}$)-plane along which 32\%,
5\%, and $4 \times 10^{-3}$ of simulations of the cumulative
distribution of 285 bursts drawn from the peak flux distribution of the
model have KS deviations $D$ larger than that of the data.  Panel (d)
compares the brightness distribution of the model shown in (a) - (c),
taking $\delta t=30$~Myrs and $d_{max}=200$~kpc, to the BATSE plus PVO
data.
}
\vspace{-5mm}
\end{figure}

\section*{Comparison between models and data}
We compare the models with a carefully-selected data set that is
self-consistent.  We use only bursts that trigger on the 1024~ms
timescale because we require that all bursts lie above the counts
threshold in one trigger timescale;  the 1024~ms timescale yields the
largest sampling depth, and therefore imposes the strongest constraint
on models, of the three BATSE trigger timescales.  We adopt $F_{\rm
pk}^{1024}$, the peak flux in 1024~ms, as our measure of burst
brightness.  We therefore include only bursts which have a $F_{\rm
pk}^{1024}$ and $t_{90} > 1024$~ms.  We consider only bursts with
$F_{\rm pk}^{1024} \ge 0.35$~photons~cm$^{-2}$~s$^{-1}$ in order to
avoid threshold effects \cite{Fenimore:93,ZandtFen:94}.  We also
exclude overwriting bursts, because the threshold is much higher for
these bursts, and MAXBC bursts, because they have unknown positional
errors.  The 2B catalogue contains 285 bursts satisfying the above
criteria.  This set of bursts has Galactic dipole and quadrupole
moments $\langle \cos \theta \rangle =0.056 \pm 0.034$, and $\langle
\sin^2 b-{1\over 3} \rangle = -0.033 \pm 0.017$, compared to the values
$\langle \cos \theta \rangle =-0.013$, and $\langle \sin^2 b-{1\over 3}
\rangle = -0.005$ expected for a uniform sky distribution, taking into
account the BATSE sky exposure.

As a first step in testing the viability of Galactic halo models, we
have compared the Galactic dipole and quadrupole moments, $\langle \cos
\theta \rangle$ and $\langle \sin^2 b - 1/3 \rangle$, of the angular
distribution of bursts for the model with those for the above set of
bursts, using $\chi^2$.  We have also compared the peak flux
distribution for the model with that for the above set of bursts, using
the KS test.

\begin{figure}[t]
\centerline{\psfig{file=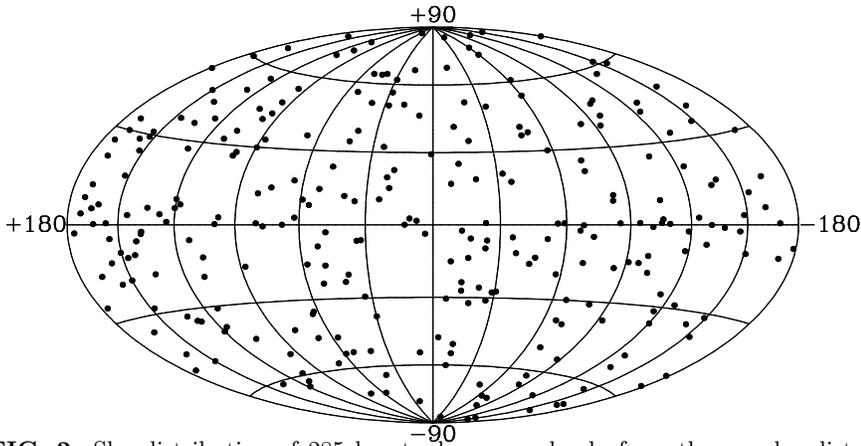,width=11cm,angle=-90}}
\caption{Sky distribution of 285 bursts drawn randomly
from the angular distribution expected for the model
illustrated in Figure~1 when $\delta t=30$~Myrs, and
$d_{\rm max}= 200$~kpc.}
\end{figure}

These comparisons do {\it not} provide estimates of model parameters
(i.e., they do not yield parameter confidence regions), but are meant
only to be a rough ``goodness-of-fit" guide to models which should be
tested using a more rigorous approach like the maximum likelihood
method.

As an illustrative example, we show in Figure~1 the results for a
Galactic halo model in which neutron stars are born with a uniform
single velocity $1000$~\kms\ and the burst-active phase has an initial
burst rate $r \propto t^2$ and lasting $\Delta t = 500$ million years.
Figure 2 shows the sky distribution of 285 bursts drawn randomly from
the angular distribution expected for the model with $\delta
t=30$~Myrs, and $d_{\rm max}= 200$~kpc.

Comparisons of this kind show that the high-velocity neutron star model
can reproduce the peak flux and angular distributions of the bursts in
the BATSE 2B catalogue for neutron star kick velocities $v_{\rm kick}
\simgreat 800$~\kms, burst turn-on ages $\delta t \simgreat
10$~million years, and BATSE sampling depths $100$~kpc $\simless d_{\rm
max} \simless 400$~kpc.  Moreover, comparisons of this kind show that
there is a large region of parameter space in which these models can
reproduce the angular distribution of the bursts in the preliminary
BATSE~3B catalogue.

In high-velocity neutron star models, the slope of the cumulative peak
flux distribution for the brightest BATSE bursts and the PVO bursts
reflects the space density of the relatively small fraction of burst
sources in the solar neighborhood.  The nearness of the observed slope
of the cumulative peak flux distribution of these bursts to -3/2, the
value expected for a uniform spatial distribution of sources which emit
bursts that are ``standard candles," must be considered a coincidence
in the high-velocity neutron star model.  However, a spread in neutron
star kick velocities, in neutron star ages at which bursting behavior
begins, or in the burst luminosity function tends to produce a
cumulative peak flux distribution with a slope of -3/2; beaming of
bursts along the direction of motion of the source or evolution of the
rate of bursting as a function of age also tends to produce a slope of
-3/2.

We find that there are many combinations of these factors which
successfully reproduce the slope of the BATSE plus PVO peak flux
distribution.  For example, a model in which the burst luminosity
function is a log normal distribution with a FWHM of a factor of
$\simless 10$ and the burst-active phase has an abrupt (``heaviside
function") turn-on, one in which the kick velocities are distributed
between $800$ and $ 1200$~km~s$^{-1}$ and the burst-active phase
initially has a burst rate $r = (t/\delta t)$, or one in which the
burst-active phase initially has a burst rate $r = {1 \over 2}
(t/\delta t)^2$ work equally well.  Figure~1d compares the peak flux
distribution for the last model and the BATSE+PVO peak flux
distribution.

M31 provides a strong constraint on the BATSE sampling distance $d_{\rm
max}$ \cite{Hak:94}.  We have investigated the effects of M31 within the
framework of the high-velocity neutron star model described above by
including the distortion of the Galactic halo potential due to M31 and
the burst sources emanating from M31.  We find that for such models M31
imposes a limit on the BATSE sampling distance $d_{\rm max} \simless
400$~kpc, even if the bursting activity of neutron stars lasts for more
than $10^9$ years \cite{BulCopLam:95c}.

%

%



\begin{references}


\bibitem {KSO:739}
R. W. Klebesadel, I. B. Strong, and R. A. Olson, Ap.J., {\bf 182}, L85
(1973).

\bibitem {Hig:Lin:90}
J. C. Higdon, and R. E. Lingenfelter 1990, Ann. Rev. Astron. Ap., {\bf
28}, 401 (1990).

\bibitem {Hard:91}
A. Harding, Phys. Reports, {\bf 206}, 327 (1991).

\bibitem {Meegan:92}
C. A. Meegan, et al., Nature, {\bf 355}, 143 (1992).

\bibitem {Briggs:95}
M. S. Briggs, et~al., Ap. J., submitted (1995).

\bibitem {Hak:94}
J. Hakkila, et~al., Ap. J., {\bf 422}, 659 (1994).

\bibitem {Hartmann:94}
D. Hartmann, et~al., Ap. J. Suppl., {\bf 90}, 893 (1994).

\bibitem {Fenimore:93}
E. E. Fenimore, et al., Nature, {\bf 366}, 40 (1993).

\bibitem {LyneLori:94}
A. G. Lyne and D. R. Lorimer, Nature, {\bf 369}, 127 (1994).

\bibitem {Frail:94}
D. A. Frail, W. M. Goss, and J. B. Whiteoak, Ap. J., {\bf 437}, 781
(1994).

\bibitem {LiDer:92}
H. Li and C. Dermer, Nature, {\bf 359}, 514 (1992).

\bibitem {Pods:94}
Ph. Podsiadlowski, M. J. Rees, and M. Ruderman, M.N.R.A.S., {\bf },  (1995).

\bibitem {KG:89}
K. Kuijken and G. Gilmore 1989, M.N.R.A.S., {\bf 239}, 571 (1989).

\bibitem {ZandtFen:94}
J. J. M. in~'t~Zand and E.~E.~Fenimore, in Gamma-Ray Bursts, AIP
Conference Proceedings No.~307, ed.~G.~J.~Fishman, J.~J.~Brainerd, and
K.~Hurley (New York: AIP), p. 692 (1994).

\bibitem {BulCopLam:95c}
T. Bulik, P. S. Coppi, and D. Q. Lamb, in preparation (1995).


\end{references}
\end{document}